\documentclass[conference]{IEEEtran}
\IEEEoverridecommandlockouts

\usepackage{cite}
\usepackage{amsmath,amssymb,amsfonts}
\usepackage{algorithmic}
\usepackage{graphicx}
\usepackage{textcomp}
\usepackage[table,xcdraw]{xcolor}
\usepackage{booktabs}
\usepackage{pifont}
\usepackage{enumitem}
\usepackage{multirow}
\usepackage{url}
\usepackage[most]{tcolorbox}

\def\BibTeX{{\rm B\kern-.05em{\sc i\kern-.025em b}\kern-.08em
    T\kern-.1667em\lower.7ex\hbox{E}\kern-.125emX}}

\newcommand{\head}[1]{\noindent \textbf{#1}}

\newcommand{\Code}[1]{\lstinline{#1}}
\usepackage{listings}  
\begin{document}

\title{Good News for Script Kiddies? Evaluating Large Language Models for Automated Exploit Generation}


\author{
    David Jin$^{1}$, Qian Fu$^{2}$, Yuekang Li$^{1}$\\
    \textit{$^{1}$UNSW Sydney, Australia, \{david.jin,yuekang.li\}@unsw.edu.au}\\
    \textit{$^{2}$Commonwealth Scientific and Industrial Research Organisation (CSIRO) Data61, Australia, qian.fu@data61.csiro.au}
}

\maketitle

\begin{abstract}
Large Language Models (LLMs) have demonstrated remarkable capabilities in code-related tasks, raising concerns about their potential for automated exploit generation (AEG). This paper presents the first systematic study on LLMs' effectiveness in AEG, evaluating both their cooperativeness and technical proficiency. To mitigate dataset bias, we introduce a benchmark with refactored versions of five software security labs. Additionally, we design an LLM-based attacker to systematically prompt LLMs for exploit generation. Our experiments reveal that GPT-4 and GPT-4o exhibit high cooperativeness, comparable to uncensored models, while Llama3 is the most resistant. However, no model successfully generates exploits for refactored labs, though GPT-4o's minimal errors highlight the potential for LLM-driven AEG advancements.
\end{abstract}

\begin{IEEEkeywords}
Exploitability, Software Vulnerability, Automated Exploit Generation, Large Language Model.
\end{IEEEkeywords}

\section{Introduction}

Software vulnerabilities pose significant threats to software security. Assessing their severity is crucial for prioritizing mitigation efforts. 
Exploitability serves as a key metric for evaluating severity~\cite{Elder2024ASO}, and automated exploit generation (AEG) enhances efficiency by reducing manual effort in exploitability evaluation~\cite{Chen2020KOOBETF}.

Research on automated exploit generation (AEG) began in the early 2000s~\cite{Heelan2009MScCS, 2008IS}, but early techniques relied on either patched programs or crashing inputs. 
In 2011, Avgerinos et al. introduced an end-to-end approach leveraging preconditioned symbolic execution~\cite{Avgerinos2011AEGAE}. 
Subsequent studies~\cite{Wu2018FUZETF, Chen2020KOOBETF, Padaryan2015AutomatedEG, Xu2018AutomaticEG, Iannone2021TowardAE} explored exploit path discovery in vulnerable programs, generating Proof-of-Concept (PoC) test cases via symbolic execution or fuzzing. 
However, these methods require significant expertise due to their reliance on complex program analyses.

The emergence of large language models (LLMs) has reduced the expertise required for various code-related tasks~\cite{Hou2023LargeLM}, including security applications such as security test generation~\cite{Zhang2023HowWD}, penetration testing~\cite{Deng2024PentestGPTEA}, and taint analysis~\cite{Zhao2024LeveragingSR}. 
However, no systematic study has examined LLMs' performance in automated exploit generation (AEG). 
This gap raises concerns about the accessibility of exploitation for novice attackers while also presenting opportunities to advance AEG research.

To address this gap, we present the first systematic study on LLMs' performance in automated exploit generation (AEG). Conducting this study requires overcoming two key challenges:
\ding{182} We use five labs from SEED Labs' software security exercises~\cite{seedlabs}. However, publicly available datasets pose a risk of bias, as LLMs may have been trained on SEED Labs. To mitigate this, we created an additional dataset by manually refactoring five labs while preserving their vulnerabilities.
\ding{183} Most LLMs, except uncensored ones, refuse exploit generation requests, requiring multi-round interactions to guide them towards the desired output. Different LLMs respond variably, making manual interactions inconsistent. To address this, we developed an LLM-based attacker that autonomously interacts with different LLMs, ensuring a systematic and reproducible AEG evaluation.


Our study aims to answer the following research questions:
\begin{itemize}[leftmargin=*]
    \item \textbf{RQ1:} To what extent do LLMs comply with exploit generation requests?
    \item \textbf{RQ2:} How effective are LLMs in generating working exploits?
\end{itemize}

Our experiments show that GPT-4 and GPT-4o exhibit a high degree of cooperation in exploit generation, comparable to some uncensored open-source models. 
Among the evaluated models, Llama3 was the most resistant to such requests.
Despite their willingness to assist, the actual threat posed by these models remains limited, as none successfully generated exploits for the five custom labs with refactored code. 
However, GPT-4o, the strongest performer in our study, typically made only one or two errors per attempt. 
This suggests significant potential for leveraging LLMs to develop advanced, generalizable AEG techniques.


In summary, this paper makes the following contributions:
\begin{itemize}[leftmargin=*]
\item \textbf{Systematic Study.} The first systematic evaluation of LLMs' performance in automated exploit generation (AEG).
\item \textbf{LLM-Based Attacker.} A novel LLM-driven attacker that systematically guides LLMs to generate exploits.
\item \textbf{Benchmark.} A curated benchmark with refactored versions of five software security labs to mitigate dataset bias.
\end{itemize}

We release our raw experiment data and code here: \url{https://anonymous.4open.science/r/AEG_LLM-EAE8/}.
\section{Background and Related Work}

\subsection{AEG Techniques.}
An exploit is a method or code that leverages a vulnerability for malicious purposes, such as information leakage or arbitrary code execution. 
Research on automated exploit generation (AEG) began in the early 2000s. 
In 2008, Brumley et al.\cite{2008IS} demonstrated that hints from patches could facilitate exploit generation. 
The following year, Heelan and Kroening\cite{Heelan2009MScCS} introduced a technique for automatically generating control flow hijacking exploits. 
Subsequent studies advanced AEG methods~\cite{Avgerinos2011AEGAE, Chen2020KOOBETF, Padaryan2015AutomatedEG, Xu2018AutomaticEG, Iannone2021TowardAE}, with Avgerinos et al.~\cite{Avgerinos2011AEGAE} pioneering the first end-to-end approach using preconditioned symbolic execution, establishing a trend toward exploit path search-based techniques.

Despite progress in AEG, existing techniques remain constrained to specific target programs (e.g., operating system kernels~\cite{Chen2020KOOBETF}) or specific vulnerability types (e.g., use-after-free vulnerabilities~\cite{Wu2018FUZETF}).
Moreover, these methods rely on complex program analysis, requiring substantial domain expertise. 
Currently, there is no general, easy-to-use AEG technique.



\subsection{LLMs for Software Security.}
By applying neural scaling laws~\cite{Bahri2021ExplainingNS} to deep learning architectures like transformers, AI researchers developed large language models (LLMs), which quickly gained prominence for their effectiveness in both natural language processing (NLP) and coding tasks~\cite{Hou2023LargeLM}. 
LLMs have been adopted in various software security applications; for instance, Deng et al.\cite{Deng2024PentestGPTEA} proposed an LLM-based penetration testing framework that enables less experienced testers to achieve competitive results. 
However, despite their potential for automating security tasks, LLMs often refuse harmful requests, such as exploit generation, due to internal alignment mechanisms\cite{Xu2024ACS}.


\subsection{Threat Model.}
In this study, we assume the target programs contain known vulnerabilities accessible to the attacker, who also has access to their source code. 
The attacker can provide the vulnerable program to an LLM, requesting automated exploit generation. 
If the LLM successfully generates exploits, the attacker can leverage them for malicious purposes, such as arbitrary code execution.


\section{Study Design}

\begin{figure}[t]
    \centering
    \includegraphics[width=\linewidth]{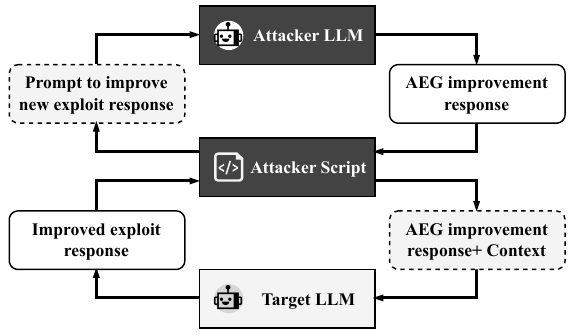}
    \caption{Workflow of the LLM-based attacker}
    \label{fig:workflow}
\end{figure}

\subsection{Benchmark Construction}
We selected five software security labs from SEED Labs~\cite{seedlabs} as benchmark vulnerable programs for evaluation. The Environment Variable and Set-UID Lab and Shellcode Development Lab were excluded due to challenges in automated verification, which we leave for future work. The chosen labs cover a range of common software vulnerabilities, including buffer overflow, return-to-libc attacks, format string vulnerabilities, race conditions, and dirty COW exploits.

To ensure an objective evaluation of LLMs, we further processed the source code. Each lab was tested in its original form without guiding comments, which are unlikely to appear in real-world code. Additionally, we created a modified version by renaming variables and functions to \Code{varX} and \Code{functionX}, where \Code{X} is assigned sequentially to obfuscate the code. This step aimed to mitigate the risk of LLMs recalling memorized solutions from training data. Consequently, our benchmark consists of two sets: the original and obfuscated labs.

To establish a reference for evaluation, we manually solved each lab. The target LLM’s final code output was then compared against these solutions. The evaluation metric was the number of errors in the LLM-generated code that prevented the exploit from functioning correctly.

\subsection{Evaluated LLMs}
We evaluated five LLMs: OpenAI’s GPT-4o and GPT-4o-mini, along with Llama3 (8B), Dolphin-Mistral (7B), and Dolphin-Phi (2.7B). These models represent a diverse selection of potential tools for attackers. Llama3 serves as a censored general-purpose model, Dolphin-Mistral as an uncensored coding-specific model, and Dolphin-Phi as an uncensored general-purpose model. The locally run models were executed via Ollama~\cite{ollama}, while OpenAI’s models provide a comparison against larger, proprietary architectures.




\subsection{LLM-Based Attacker}
Prompting LLMs to generate exploits often encounters resistance, requiring multi-turn interactions to achieve results. However, different LLMs respond unpredictably, making the conversation dynamic and non-deterministic. While this process can be conducted manually, it contradicts the automated nature of AEG and introduces evaluator bias based on an attacker's experience and expertise. To address this, we designed an automated attacker program that leverages an LLM to systematically instruct target LLMs under evaluation.

Figure~\ref{fig:workflow} illustrates the workflow of the LLM-based attacker. In our implementation, OpenAI’s GPT-4o was selected as the attacker model due to its strong reasoning capabilities at the time of experimentation. The model generating exploits is referred to as the target LLM. An attacker script mediates communication between the attacker LLM and the target LLM by passing prompts and responses. The script initiates an API request to the target LLM for an initial attempt at solving a lab. This response is then provided to the attacker LLM, which generates an improved prompt instructing the target LLM on refining its exploit attempt. The process repeats iteratively, with the improved response sent back to the target LLM, and its output further refined by GPT-4o-mini. This cycle continues for up to 15 iterations (empirically determined) or until the attacker LLM determines that no further improvements can be made.

\section{Results and Analysis}

\subsection{RQ1: Cooperativeness}

\begin{table*}[ht]
\caption{Average cooperation percentage of LLMs during iterative improvement process}
\centering
\begin{tabular}{llrrrrr|r}
\hline
                                                             &                                      & \multicolumn{1}{l}{Buffer overflow} & \multicolumn{1}{l}{Return to libc} & \multicolumn{1}{l}{Format string} & \multicolumn{1}{l}{Race condition} & \multicolumn{1}{l|}{Dirty COW} & \multicolumn{1}{l}{Average} \\ \hline
\multicolumn{1}{l|}{\multirow{2}{*}{Closed-Source}}          & \multicolumn{1}{l|}{GPT-4o}          & 100                                 & 100                                & 100                               & 92                                 & 93                             & 97                          \\ \cline{2-8} 
\multicolumn{1}{l|}{}                                        & \multicolumn{1}{l|}{GPT-4o-mini}     & 100                                 & 92                                 & 100                               & 88                                 & 100                            & 96                          \\ \hline
\multicolumn{1}{l|}{Open-Source Censored}                    & \multicolumn{1}{l|}{Llama3}          & 27                                  & 13                                 & 64                                & 16                                 & 17                             & \cellcolor[HTML]{90EE90}{27}                           \\ \hline
\multicolumn{1}{l|}{\multirow{2}{*}{Open-Source Uncensored}} & \multicolumn{1}{l|}{Dolphin Mistral} & 100                                 & 100                                & 100                               & 91                                 & 75                             & 93                          \\ \cline{2-8} 
\multicolumn{1}{l|}{}                                        & \multicolumn{1}{l|}{Dolphin Phi}     & 97                                  & 100                                & 100                               & 84                                 & 92                             & 95                          \\ \hline
\multicolumn{2}{l|}{Average}                                                                        & 85                                  & 81                                 & 93                                & \cellcolor[HTML]{90EE90}{74}                                 & \cellcolor[HTML]{90EE90}{75}                             & \multicolumn{1}{l}{}        \\ \hline
\end{tabular}
\label{table:cooperation}
\end{table*}

Table~\ref{table:cooperation} shows the average percentage of the cooperative response from the target LLMs during the iterative exploit improvement process when interacting with the LLM-based attacker.
The higher the value is, the more \textit{willing} an LLM is to help with exploit generation.
We can observe that Llama3 is the least cooperative model.
Surprisingly, we can also find that the GPT-series models are quite willing to aid the user with the exploit generation tasks, demonstrating a similar level of cooperativeness to uncensored models.
As for types of vulnerabilities, LLMs are more likely to refuse the requests for exploiting race condition and dirty COW vulnerabilities.
We are led to speculate that these topics are more likely to be censored.

\begin{tcolorbox}[colback=black!6!white,enhanced,frame hidden, boxsep=0pt,left=5pt,right=5pt,top=2pt,bottom=2pt]
\textbf{Answer to RQ1:}
Due to the alignment of LLMs, they may not cooperate when requested to generate exploits.
The cooperativeness is affected by both the model types and the vulnerability types.
The OpenAI models are surprisingly cooperative to help with exploit generation.
\end{tcolorbox}


\subsection{RQ2: Effectiveness}

\begin{figure}[t]
    \centering
    \includegraphics[width=\linewidth]{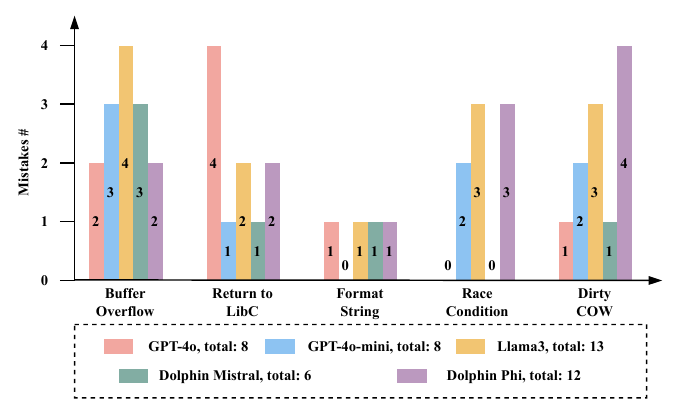}
    \caption{Number of mistakes made by LLMs on original SEED Lab programs}
    \label{fig:orig}
\end{figure}

\begin{figure}[t]
    \centering
    \includegraphics[width=\linewidth]{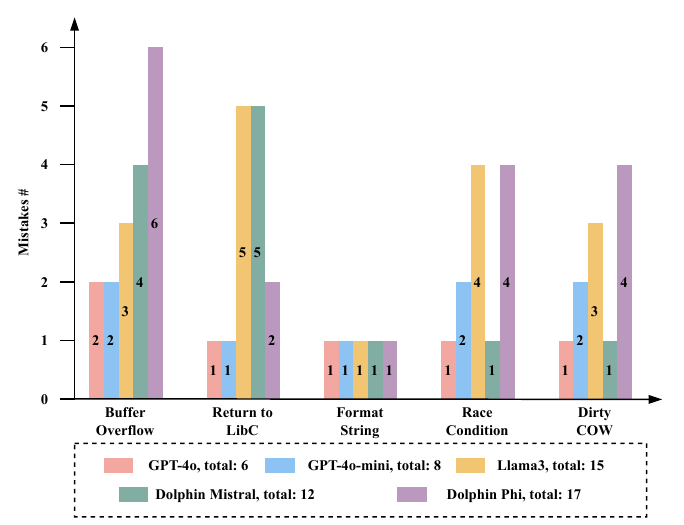}
    \caption{Number of mistakes made by LLMs on refactored SEED Lab programs}
    \label{fig:refactor}
\end{figure}

Figure~\ref{fig:orig} shows the number of mistakes in the exploits created by each target LLM on each original lab program.
Figure~\ref{fig:refactor} shows the number of mistakes made by each target LLM on every refactored lab program.
From the two figures, we can observe that the GPT-series of models are the best performing models in terms of the exploit quality.
An interesting thing is that Dolphin Mistral performs the best with the original labs while it made significantly more mistakes on the refactored programs.
This indicates that Dolphin Mistral might have been trained with SEED Lab materials.
We can also observe for a performance drop for all the evaluated LLMs on the refactored programs, as some LLMs may successfully generate exploits for the format string vulnerability and the race condition vulnerability with the original programs while none of them can succeed on the refactored programs.
This indicates either the importance of code symbols for AEG tasks with LLMs or that some LLMs might have been trained with the SEED Labs materials.

\begin{tcolorbox}[colback=black!6!white,enhanced,frame hidden, boxsep=0pt,left=5pt,right=5pt,top=2pt,bottom=2pt]
\textbf{Answer to RQ2:}
The GPT-series of models are the best performing ones for AEG tasks yet no evaluated LLMs can succeed on the refactored programs.
Although AEG with LLMs is a promising direction, more research effort is needed to improve the performance.
\end{tcolorbox}



\subsection{Detailed Analysis}

Here we identify common errors made across most AEG attempts, their rates of occurrence among all relevant attempts and provide some speculation on why these errors occur. 

\head{Buffer Overflow.} The most common error made is failing to create a NOP sled of sufficient length. This was observed in 60\% of attempts made. Since the NOP sled length is derived from the size of an unprotected read instruction, it is believed that the AEG LLM never forms the connection between an unprotected read and a buffer overflow attack. Thus, it fails to create a NOP sled of sufficient length to take advantage of the unsafe read instruction.

A less common error, occuring in 30\% of AEG attempts, is having the incorrect order of the shellcode and the return address. While most attempts were successful in this regard, it is notable that 3 out of 10 attempts suffered this mistake. It suggests a failing to make a connection between the return address and payload, specifically that having the return address allows program execution until the payload is encountered.

While not strictly an error, in 50\% of cases, the buffer was first overflowed with junk data and then a NOP sled was constructed. In practice using only NOP characters for this process suffices. It only serves to complicate the program when junk characters are used that are not NOPs.

\head{Return to LibC.} An error that pervades 100\% of all AEG attempts is an incorrect padding size in the payload. This padding size is calculated by the difference between the buffer and ebp address. Therefore, one possible reason for this mistake is a failing of LLMs in performing arithmetic accurately. Alternatively, the LLMs may not have made the connection between the padding size and the aforementioned difference. 

Another common error is the use of an incorrect, or absence of, an address for the system, exit, and shell commands. These are vital for the payload to function as they must be called successively. This error was present in 50\% of test cases. There is possibly an effect akin to losing track of the goal as the LLMs can produce a result that looks like a valid payload but ultimately contains junk data where these addresses are meant to be. Thus, they may understand the shape of the solution but cannot make use of the given addresses as they fail to understand the goal of the payload.

\head{Format String.} In the format string tests, 90\% of attempts failed to produce an output that would cause the program to crash. Since an actual string that does so is rather simple, it is hard to ascribe to the LLMs any one common mistake. However, in general, outputs are much longer than necessary, resulting in what appear to be over-engineered solutions.

\head{Race Condition.} Many outputs failed to loop the proposed attack which, given that it is probabilistic in nature, means it is unlikely to work in practice. This particular error occurred in 60\% of test cases. It suggests a failure to understand that looping the attack is necessary outside of a theoretical context, potentially indicating a gap between theory and practice in LLM training data.

Another error was a failure to remove the symbolic link in order to reset for the next attempt. This also occurred in 60\% of cases. While resetting is not necessary if one is not looping the attack, it should be noted that the attempts that failed to remove the symbolic link are not exactly those that failed to loop the attack. In particular Dolphin-Mistral fails to loop but removes the symbolic link on the renamed attempt, while 4o fails to remove the link but successfully loops the attack. This suggests that LLMs fail to make the connection between looping attacks and what must be reset for those attacks.

\head{Dirty COW.} In a similar vein to the race condition attack, many attempts failed to loop this attack despite its probabilistic nature. This was seen in 60\% of test cases and is suggestive of issues already discussed in the race condition evaluation.

Another error in Dirty COW attempts was the use of mutexes or semaphores which prevent race conditions. This was present in 60\% of cases. While this might be desirable in some contexts, this exploit relies on a race condition vulnerability to be successful. This suggests that LLMs do not identify a link between this kind of attack and the purpose of mutexes or semaphores. It is possible that LLMs have seen a pattern of such features in code being associated with the concept of good quality programming. In this case, instead of identifying links between when they are desired or not, they have simply understood that it must be desirable in all contexts because it shows up statistically more often around favourable keywords.

It appears that a general trend in the data is that current LLMs understand what an attack for a common vulnerability will ‘look’ like without having formed connections about the underlying mechanisms. However, the above is purely speculative, and the postulated theories have not been rigorously tested.

\section{Limitations and Future Work}

This study provides a preliminary investigation into the use of LLMs for automated exploit generation (AEG). While our findings offer insights into this emerging research direction, several limitations remain. Below, we discuss these limitations and outline future work to enhance this study.

\head{Reasoning Models.}
During our research, OpenAI and DeepSeek released reasoning-focused models (GPT-o1 and DeepSeek-r1), which demonstrate stronger performance in logic-intensive tasks. These models may be better suited for AEG, and we plan to evaluate them in future studies to obtain more comprehensive insights.

\head{Real-World Exploits.}
We used SEED Labs due to their well-documented vulnerabilities and carefully designed programs, making them a reliable ground-truth dataset. To improve benchmark diversity, future work will incorporate real-world vulnerabilities and exploits.
\section{Conclusion}
Our study provides the first systematic evaluation of LLMs in automated exploit generation (AEG), revealing that models like GPT-4 and GPT-4o exhibit high cooperativeness, comparable to some uncensored open-source models, while Llama3 is the most resistant. Despite their willingness to assist, none of the evaluated models successfully generated exploits for refactored vulnerabilities, though GPT-4o demonstrated minimal errors. These findings suggest that while current LLMs pose limited immediate risk for exploit generation, their evolving capabilities present significant potential for advancing AEG techniques. Our benchmark and LLM-based attacker offer a foundation for future research in this domain.

\bibliographystyle{IEEEtran}
\bibliography{ref}

\begin{thebibliography}{10}
\providecommand{\url}[1]{#1}
\csname url@samestyle\endcsname
\providecommand{\newblock}{\relax}
\providecommand{\bibinfo}[2]{#2}
\providecommand{\BIBentrySTDinterwordspacing}{\spaceskip=0pt\relax}
\providecommand{\BIBentryALTinterwordstretchfactor}{4}
\providecommand{\BIBentryALTinterwordspacing}{\spaceskip=\fontdimen2\font plus
\BIBentryALTinterwordstretchfactor\fontdimen3\font minus \fontdimen4\font\relax}
\providecommand{\BIBforeignlanguage}[2]{{%
\expandafter\ifx\csname l@#1\endcsname\relax
\typeout{** WARNING: IEEEtran.bst: No hyphenation pattern has been}%
\typeout{** loaded for the language `#1'. Using the pattern for}%
\typeout{** the default language instead.}%
\else
\language=\csname l@#1\endcsname
\fi
#2}}
\providecommand{\BIBdecl}{\relax}
\BIBdecl

\bibitem{Elder2024ASO}
\BIBentryALTinterwordspacing
S.~Elder, M.~R. Rahman, G.~Fringer, K.~Kapoor, and L.~Williams, ``A survey on software vulnerability exploitability assessment,'' \emph{ACM Computing Surveys}, vol.~56, pp. 1 -- 41, 2024. [Online]. Available: \url{https://api.semanticscholar.org/CorpusID:268570392}
\BIBentrySTDinterwordspacing

\bibitem{Chen2020KOOBETF}
\BIBentryALTinterwordspacing
W.~Chen, X.~Zou, G.~Li, and Z.~Qian, ``Koobe: Towards facilitating exploit generation of kernel out-of-bounds write vulnerabilities,'' in \emph{USENIX Security Symposium}, 2020. [Online]. Available: \url{https://api.semanticscholar.org/CorpusID:209515033}
\BIBentrySTDinterwordspacing

\bibitem{Heelan2009MScCS}
\BIBentryALTinterwordspacing
S.~Heelan and D.~Kroening, ``Automatic generation of control flow hijacking exploits for software vulnerabilities,'' 2009. [Online]. Available: \url{https://api.semanticscholar.org/CorpusID:62247176}
\BIBentrySTDinterwordspacing

\bibitem{2008IS}
\BIBentryALTinterwordspacing
``Automatic patch-based exploit generation is possible: Techniques and implications.'' [Online]. Available: \url{https://api.semanticscholar.org/CorpusID:8081699}
\BIBentrySTDinterwordspacing

\bibitem{Avgerinos2011AEGAE}
\BIBentryALTinterwordspacing
T.~Avgerinos, S.~K. Cha, B.~L.~T. Hao, and D.~Brumley, ``Aeg: Automatic exploit generation,'' in \emph{Network and Distributed System Security Symposium}, 2011. [Online]. Available: \url{https://api.semanticscholar.org/CorpusID:14420062}
\BIBentrySTDinterwordspacing

\bibitem{Wu2018FUZETF}
\BIBentryALTinterwordspacing
W.~Wu, Y.~Chen, J.~Xu, X.~Xing, X.~Gong, and W.~Zou, ``Fuze: Towards facilitating exploit generation for kernel use-after-free vulnerabilities,'' in \emph{USENIX Security Symposium}, 2018. [Online]. Available: \url{https://api.semanticscholar.org/CorpusID:52052904}
\BIBentrySTDinterwordspacing

\bibitem{Padaryan2015AutomatedEG}
\BIBentryALTinterwordspacing
V.~A. Padaryan, V.~V. Kaushan, and A.~N. Fedotov, ``Automated exploit generation for stack buffer overflow vulnerabilities,'' \emph{Programming and Computer Software}, vol.~41, pp. 373 -- 380, 2015. [Online]. Available: \url{https://api.semanticscholar.org/CorpusID:11233774}
\BIBentrySTDinterwordspacing

\bibitem{Xu2018AutomaticEG}
\BIBentryALTinterwordspacing
L.~Xu, W.~Jia, W.~Dong, and Y.~Li, ``Automatic exploit generation for buffer overflow vulnerabilities,'' \emph{2018 IEEE International Conference on Software Quality, Reliability and Security Companion (QRS-C)}, pp. 463--468, 2018. [Online]. Available: \url{https://api.semanticscholar.org/CorpusID:52003316}
\BIBentrySTDinterwordspacing

\bibitem{Iannone2021TowardAE}
\BIBentryALTinterwordspacing
E.~Iannone, D.~D. Nucci, A.~Sabetta, and A.~D. Lucia, ``Toward automated exploit generation for known vulnerabilities in open-source libraries,'' \emph{2021 IEEE/ACM 29th International Conference on Program Comprehension (ICPC)}, pp. 396--400, 2021. [Online]. Available: \url{https://api.semanticscholar.org/CorpusID:233207774}
\BIBentrySTDinterwordspacing

\bibitem{Hou2023LargeLM}
\BIBentryALTinterwordspacing
X.~Hou, Y.~Zhao, Y.~Liu, Z.~Yang, K.~Wang, L.~Li, X.~Luo, D.~Lo, J.~C. Grundy, and H.~Wang, ``Large language models for software engineering: A systematic literature review,'' \emph{ACM Trans. Softw. Eng. Methodol.}, vol.~33, pp. 220:1--220:79, 2023. [Online]. Available: \url{https://api.semanticscholar.org/CorpusID:261048648}
\BIBentrySTDinterwordspacing

\bibitem{Zhang2023HowWD}
\BIBentryALTinterwordspacing
Y.~Zhang, W.-K. Song, Z.~Ji, D.~D. Yao, and N.~Meng, ``How well does llm generate security tests?'' \emph{ArXiv}, vol. abs/2310.00710, 2023. [Online]. Available: \url{https://api.semanticscholar.org/CorpusID:263334479}
\BIBentrySTDinterwordspacing

\bibitem{Deng2024PentestGPTEA}
\BIBentryALTinterwordspacing
G.~Deng, Y.~Liu, V.~M. Vilches, P.~Liu, Y.~Li, Y.~Xu, M.~Pinzger, S.~Rass, T.~Zhang, and Y.~Liu, ``Pentestgpt: Evaluating and harnessing large language models for automated penetration testing,'' in \emph{USENIX Security Symposium}, 2024. [Online]. Available: \url{https://api.semanticscholar.org/CorpusID:271325673}
\BIBentrySTDinterwordspacing

\bibitem{Zhao2024LeveragingSR}
\BIBentryALTinterwordspacing
J.~Zhao, Y.~Li, Y.~Zou, Z.~Liang, Y.~Xiao, Y.~Li, B.~Peng, N.~Zhong, X.~Wang, W.~Wang, and W.~Huo, ``Leveraging semantic relations in code and data to enhance taint analysis of embedded systems,'' in \emph{USENIX Security Symposium}, 2024. [Online]. Available: \url{https://api.semanticscholar.org/CorpusID:271325586}
\BIBentrySTDinterwordspacing

\bibitem{seedlabs}
\BIBentryALTinterwordspacing
W.~Du, ``Seed labs,'' 2025. [Online]. Available: \url{https://seedsecuritylabs.org/}
\BIBentrySTDinterwordspacing

\bibitem{Bahri2021ExplainingNS}
\BIBentryALTinterwordspacing
Y.~Bahri, E.~Dyer, J.~Kaplan, J.~Lee, and U.~Sharma, ``Explaining neural scaling laws,'' \emph{Proceedings of the National Academy of Sciences of the United States of America}, vol. 121, 2021. [Online]. Available: \url{https://api.semanticscholar.org/CorpusID:231918701}
\BIBentrySTDinterwordspacing

\bibitem{Xu2024ACS}
\BIBentryALTinterwordspacing
Z.~Xu, Y.~Liu, G.~Deng, Y.~Li, and S.~Picek, ``A comprehensive study of jailbreak attack versus defense for large language models,'' in \emph{Annual Meeting of the Association for Computational Linguistics}, 2024. [Online]. Available: \url{https://api.semanticscholar.org/CorpusID:267770234}
\BIBentrySTDinterwordspacing

\bibitem{ollama}
\BIBentryALTinterwordspacing
Ollama, ``Get up and running with large language models.'' 2025. [Online]. Available: \url{https://ollama.com/}
\BIBentrySTDinterwordspacing

\end{thebibliography}

\end{document}